\documentclass[prb,twocolumn, showpacs, amsmath, amssymb,superscriptaddress]{revtex4-1}

\usepackage{amssymb,amsfonts,amsmath} 
\usepackage{graphicx}
\usepackage{epsfig}
\usepackage{epstopdf}
\usepackage{color}
\usepackage{hyperref}
\usepackage{longtable}
\usepackage{bbm}
\usepackage{ulem}
\begin{document} 

\title{Electromagnetic Response during a Quench Dynamics to  Superconducting State: Time-Dependent Ginzburg-Landau Analysis}

\author{D. M. Kennes}
\affiliation{Department of Physics, Columbia University, New York, NY 10027, USA}
\author{A. J. Millis}
\affiliation{Department of Physics, Columbia University, New York, NY 10027, USA}

\begin{abstract}
We use a numerical solution of the deterministic TDGL equations to determine the response induced by a probe field in a  material quenched into a superconducting state. We characterize differences in response according to whether the probe is applied before, during, or after the phase stiffness has built up to its final steady state value. We put an emphasis on the extend to which superfluid response requires a non-negligible phase stiffness, which for the considered quench has to build up dynamically. A key finding is that the time dependent phase stiffness controls the likelihood of phase slips as well as the magnitude of the electromagnetic response. Additionally, we address the electromagnetic response expected if the probe itself is strong enough to activate phase slip processes. If the probe is applied before phase stiffness is sufficiently build up we find that phase slips occur so that the vector potential is compensated and no long term supercurrent is induced, while if applied at sufficient phase stiffness a weak probe pulse will induce a state with a long-lived supercurrent. If the probe is strong enough to activate the phase slip process the supercurrent state will only be metastable with a lifetime that scales logarithmically with the amplitude of fluctuations in the magnitude of the order parameter. Finally, we study the response to experimentally motivated probe fields (electric field that integrates to zero). Interestingly, depending on the relative time difference of the probe field to the build up of superconductivity, long-lived supercurrents can be induced even though the net change in vector potential is zero. 
\end{abstract}

\date{\today} 
\maketitle
\section{Introduction}
Quench dynamics, in other words the response of a system to sudden changes in parameters, is of great current interest in the context of cold atomic gases\cite{Bloch08} and correlated electron systems.\cite{Gogolin16,Essler16,Vidmar16} Quench dynamics may be studied experimentally by measuring the  response to a weak applied probe field.  One expects different responses according to whether the probe is applied prior to the quench, during the quench, or long enough after that the system has relaxed into a thermal or prethermal state. 

Recent reports of optically induced high transition temperature superconductivity\cite{mankowsky2015coherent,mitrano2016possible} along with theoretical interpretations\cite{Komnik16,Sentef16,Knap16,Kennes16,Babadi17,Babadi17,Sentef17,Nava17}  give a particular topicality to the question of the electromagnetic response of a material quenched into a superconducting state. In Ref.~\onlinecite{Kennes17} we studied this question within a BCS model that allowed for a time dependent magnitude of the order parameter but assumed perfect phase stiffness at all times. This analysis provided a reasonable account of the transient response at energy scales of the order of the superconducting gap or for probe fields applied a reasonable time after the quench.  However, the low frequency response is controlled by the behavior of the phase of the superconducting order parameter. The importance of the phase stiffness and of phase slips can be seen from the general expression for the supercurrent $j=\nabla \phi-2eA$ in terms of the gradient of the superconducting phase $\phi$ and the vector potential $A$. If a superconductor is quenched in the presence of a vector potential then the phase adjusts (to the extent possible) so that $\nabla \phi=2eA$ and the supercurrent is small, whereas if the vector potential is applied long after the superconductor is quenched then the phase is fixed, typically such that $\nabla \phi=0$, and the supercurrent is proportional to $-2eA$. The key issues of  the timing of the probe relative to the establishment of phase rigidity, and the strength of the probe relative to the field required to drive phase slips were beyond the scope of this previous work.

In this paper we investigate the interplay between the timing of the quench and the application of a probe field via solutions of the time-dependent Ginzburg-Landau (TDGL) equations.\cite{Cyrot73,Gorkov75,Ivlev84} While  TDGL equations have been extensively studied,\cite{Cyrot73,Gorkov75,Ivlev84}  this particular issue  seems not to have been previously considered. We concentrate mainly on the case of a one-dimensional system with periodic boundary conditions, but present a few results for the two dimensional case. The main conclusions we draw are not crucially affected by the dimensionality or geometry of the system considered but we note and discuss those aspects that are particular to the one dimensional case.   We thus use the one-dimensional case solely due to reasons of numerical convenience. We study the fully determininstic TDGL equation. The inclusion of  noise (full model A dynamics)  will be seen not to significantly affect the phenomena of interest here (see Refs.~ \onlinecite{Kobayashi16a,Kobayashi16b} for a recent interesting study of quenches using model A dynamics).

The rest of this paper is organized as follows. In section ~\ref{sec:Formalism} we present the equations to be solved and the methods of solution. In section ~\ref{sec:quench} we describe those aspects of the physics of a quench that are relevant to our analysis in sections ~\ref{sec:response} and ~\ref{sec:PhysicalPulse} of the response to differently tailored probe fields. Section ~\ref{sec:conclusion} is a summary and conclusion. An Appendix gives details of our numerical procedure and verifies convergence.

\section{Formalism\label{sec:Formalism}}
We use the deterministic  TDGL equations\cite{Gorkov75,Ivlev84} to describe the dynamics of the complex superconducting order parameter $\Delta=\left|\Delta\right| e^{i\phi}$, along with the charge density $\rho$ and current density $\vec{j}$ in the presence of electromagnetic fields represented by the vector potential $\vec{A}(t)$ and scalar potential $\Theta(t)$.  We choose units such that $\hbar=c=e=1$, implying that the superconducting flux quantum $\Phi_0\equiv hc/2e=\pi$. The equations are
\begin{align}
&\frac{1}{ D}\left(\partial_t+2i\Psi\right)\Delta=\frac{1}{\xi^{2}\beta}\Delta\left[r(t)-\beta |\Delta|^2\right]\notag\\
&\phantom{\frac{1}{ D}\left(\partial_t+2i\Psi\right)\Delta=}+\left[\vec{\nabla} -  2i\vec{A}(t)\right]^2 \Delta
\label{eq:TDGL}
\\
&\rho=\frac{\Psi-\Theta}{4\pi\lambda_{\rm TF}^2}
\label{eq:rho}
\\
&j=\sigma\left(-\nabla \Psi-\partial_t A(t)\right)+\frac{\sigma}{\tau_s}{\rm Re}\left[\Delta^*\left(\frac{\nabla}{i}-2A\right)\Delta\right] 
\label{eq:j}
\end{align}
Here $D$ is the normal state diffusion constant, $\Psi$ is the electrochemical potential per electron charge,  $\xi=\sqrt{6 D\tau_s}$  is related to the superconducting coherence length $\xi_0=\xi/\sqrt{r/\beta}$, where $\tau_s$ is the spin-flip scattering time, $\lambda_{\rm TF}$ is the Thomas-Fermi static charge screening length and  $\beta$ is a system dependent constant that sets the magnitude of the order parameter. The quench is specified by the time dependence of the parameter  $r(t)\sim[T_c(t)-T]$, with $T$ and $T_c$ the temperature and the superconducting critical temperature. We consider an ``interaction quench'' in which the system Hamiltonian is changed in such a way as to vary the transition temperature from less than to greater than the physical temperature. For definiteness we  measure lengths in units of $\xi$ and time in units of $\xi^2/D$ (which we write simply as $D^{-1}$ since $\xi$ is our unit of length)  and choose parameters $\beta=1$, $\sigma=1 $, $\tau_sD=\frac{1}{6}$  and $\lambda_{\rm TF}^2/\xi^2=1$. We chose those units for definiteness, but we verified that none of the general conclusions depend on this choice of parameters unless otherwise stated. 

The TDGL equations must be supplemented by the continuity equation  
\begin{equation}
\partial_t \rho+\vec{\nabla}\cdot\vec{j} =0
\label{eq:continuity}
\end{equation}
and  the Poisson equation for the scalar potential 
\begin{equation}
\nabla^2\Theta=-4\pi\rho.\label{eq:Poisson}
\end{equation}

We solve the coupled partial differential equations Eqs.~\eqref{eq:TDGL}-\eqref{eq:Poisson} with periodic boundary conditions in one or two dimensions using a finite difference approach. We discretized time in steps of $D\Delta t =0.001$ and space in units $\Delta x/\xi=1$ and checked numerically that the results obtained are converged with respect to $\Delta t$ and $\Delta x$ on the scale of the plots shown (see Appendix ~\ref{Appendix:a}). 

The partial differential equations require initial conditions. We assume that for $t<0$ the parameter $r<0$ so there is no superconductivity, and that at $t=0$ $r$ is suddenly switched to a positive value (we chose $r=0.1$, so $\xi_0=1/\sqrt{0.1}$). The pre-quench state at $t \leq 0$ is characterized by small thermal fluctuations, which for positive $r$ will grow exponentially, leading eventually to an equilibrium superconducting state. We therefore choose as initial condition at $t=0$ a state with random, small order parameter values. We considered two cases: (i) absolute values of the order parameter drawn randomly from a uniform distribution $|\Delta|\in [0,\Delta_{\rm ini}]$ and phase values drawn randomly from a uniform distribution $\phi\in(-\pi,\pi]$ on the different lattice sites (see Appendix ~\ref{Appendix:a} left panel of Fig.~\ref{fig:ini_cond_1}) and (ii) fixed magnitude  $|\Delta|=\Delta_{\rm ini}$ with phase values randomly drawn from a uniform distribution $\phi\in(-\pi,\pi]$ (see Appendix     ~\ref{Appendix:a} right panel of Fig.~\ref{fig:ini_cond_1}). Both initial conditions give very similar results (see Appendix ~\ref{Appendix:a})  and all of the results in the following are obtained by assuming that the initial conditions are characterized by a fixed small magnitude and a random phase.

\section{Quench Dynamics \label{sec:quench}}

\begin{figure*}
\begin{center} 
\includegraphics[width=2\columnwidth, angle=-0]{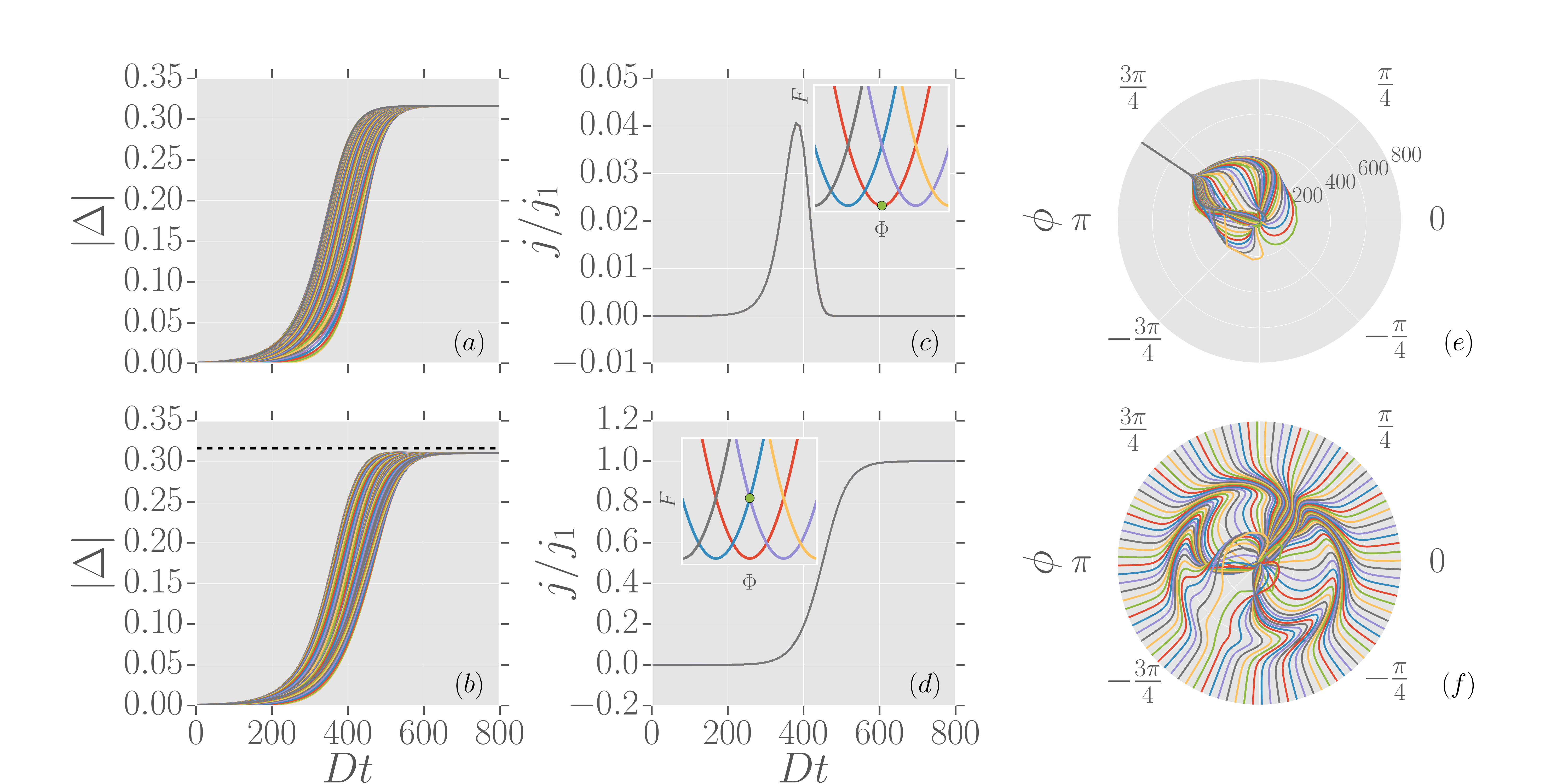}
\caption{Time evolution of order parameter magnitude (panels (a) and (b)), current (panels (c) and (d)) and order parameter phase (panels (e) and (f)) computed for a ring of size $L=100\xi$   by time-evolving an  initial configuration specified by random  phases on each site and a constant magnitude of the order parameter $\left|\Delta_{\rm ini}\right|=10^{-3}$. We show each of the $100$ different values of magnitude, current and phase at the different sites as lines of different colors. The top row (panels (a), (c), (e)) shows results obtained from an  initial condition that happens to lead to a long time state with zero current; the bottom row (panels (b), (d), (f)) shows results obtained from an   initial condition that happens to lead to to a long time state with nonzero phase winding (in this example $\nabla \phi=2\pi/L$). In panels (e) and (f) the phase is represented in  a polar coordinate system with the value of the phase as the angular coordinate and time as the radial coordinate.  The insets show the free energies as a function of the flux threading the ring. Different parabolas correspond to different integer values of the phase twist. The dots in the insets show the respective state found at large times for the respective initial conditions. In the lower panel the actual long-time state is characterized by $n=1$ so it on the parabola displaced from the $n=0$ parabola. }  
\label{fig:plot1}
\end{center}
\end{figure*}

In this section we recapitulate basic aspects of the quench dynamics in the absence of applied probe fields, in order  to set the stage for the subsequent discussion of the response to probe fields.  We consider a quench into a superconducting state achieved by instantaneously changing the interactions so that at time $t<0$ the transition temperature $T_c$ is less than the sample temperature $T$ while at time $t>0$ the transition temperature is greater. An example of two experiments for which such a study of a quench might be relevant
are given in Refs.~\onlinecite{mankowsky2015coherent,mitrano2016possible}   We suppose that the quench occurs in the presence of a spatially uniform, time independent vector potential $\vec{A}$. The non-superconducting state is characterized by small fluctuations in the superconducting order parameter. After the quench these fluctuations grow and at sufficiently long times the system evolves to a homogeneous superconducting state.   The system of main interest here is one dimensional, with periodic boundary conditions for numerical convenience. This system has the topology of a ring and an applied vector potential corresponds to a flux $\Phi= L A$ threading the ring. It is convenient to measure the vector potential in units of the superconducting flux quantum $\Phi_0=hc/2e=\pi$ (with the last equality following from the convention $\hbar=c=e=1$), writing $A=\Phi/L$ and the gauge-invariant gradient as $\vec{\nabla}-2\pi i \Phi/\left(L\Phi_0\right)$.

The possible  homogeneous superconducting states are characterized by a phase winding number $n=\oint\nabla \phi/(2\pi)$ implying a nonzero phase gradient $d\phi/dx=2\pi n/L$. The magnitude of the order parameter is\cite{Ivlev84,Kramer85} $\left|\Delta(n)\right|=\sqrt{\frac{r}{\beta}-\left( \frac{2\pi\xi}{L}\right)^2\left(n-\frac{\Phi~}{\Phi_0}\right)^2}$. The  corresponding free energy gain is $-\beta|\Delta(n)|^4$ so the true ground state will have winding number $n$ given by $2\pi$ times  the nearest integer to $\Phi/\Phi_0$  but depending on initial conditions and subsequent dynamics the actual state reached may be a metastable state with different winding number.  From Equation ~\ref{eq:j} we find that in the equilibrium state at winding number $n$ the supercurrent circulating around the ring is 
\begin{equation}
\vec{j}_n=\frac{2\pi \sigma \tau_s}{L} \left( n-\frac{\Phi~}{\Phi_0}\right)\left|\Delta(n)\right|^2. \label{eq:curr_wind}
\end{equation}

Changes in the winding number occur via phase slip processes at which the amplitude of the order parameter is driven locally to zero and the phase difference across the region with locally zero order parameter changes by a multiple of $2\pi$.\cite{Vodolazov02,Yu08,Ludac08,Ludac09,Michotte04,Baranov11}  In the deterministic dynamics studied here phase slips occur when the  local amplitude of the current is greater than an energy barrier determined by the local magnitude of the order parameter. (Note that the random initial conditions mean that this magnitude will be different on different sites and that  the random currents implied by the random phases will lead to different order parameter magnitudes at intermediate times even if the initial condition is a space independent order parameter magnitude).  Phase slips may occur as the system equilibrates, and  will be more common soon  after the quench when the order parameter is small and the energy barrier to phase slips is less.   In model A stochastic dynamics there will be a small amplitude for phase slips even if the drive is not large enough to overcome the energy barrier; for small noise the resulting corrections are exponentially small in the reciprocal of the noise amplitude and will not be considered here. 

 Fig.~\ref{fig:plot1} shows two examples of the evolution of the system following a quench with $A=0$.  The left column shows the time evolution of the order parameter magnitude  on the $100$ different sites in the system (for $L=100\xi)$, computed from two different randomly chosen initial conditions.  The initial stages of the growth are exponential with the differences between different sites arising from the randomness in the initial conditions.  The middle panels show the current on the $100$ sites. The currents are almost identical on every site because the charge fluctuations decay away almost instantaneously, leading to a state with $\vec{\nabla}\cdot\vec{j} =0$. 
 
From panel (c) we see that at early times the current is very small because the magnitude of the order parameter is very small; at intermediate times we see a current pulse associated with the equilibration of the phase degrees of freedom while the current vanishes at long times because the long-time limit of the winding number $n=0$. Panel (e) shows that after an intially complicated evolution the phase locks into a common value at all sites, corresponding to the zero current state shown in panel (c).  Panel (d) shows that for a different choice of initial conditions the long-time limit corresponds to a non-vanishing current. In this case, at  large times the phase is stiff and the magnitude of the order parameter is sufficiently large that the induced supercurrent is not large enough to drive a phase slip. Panel (f) shows the phase in the polar plot described above. The non-vanishing phase gradient is revealed as a phase monotonically increasing as one moves around the ring.  

\begin{figure}
\begin{center} 
\includegraphics[width=\columnwidth, angle=-0]{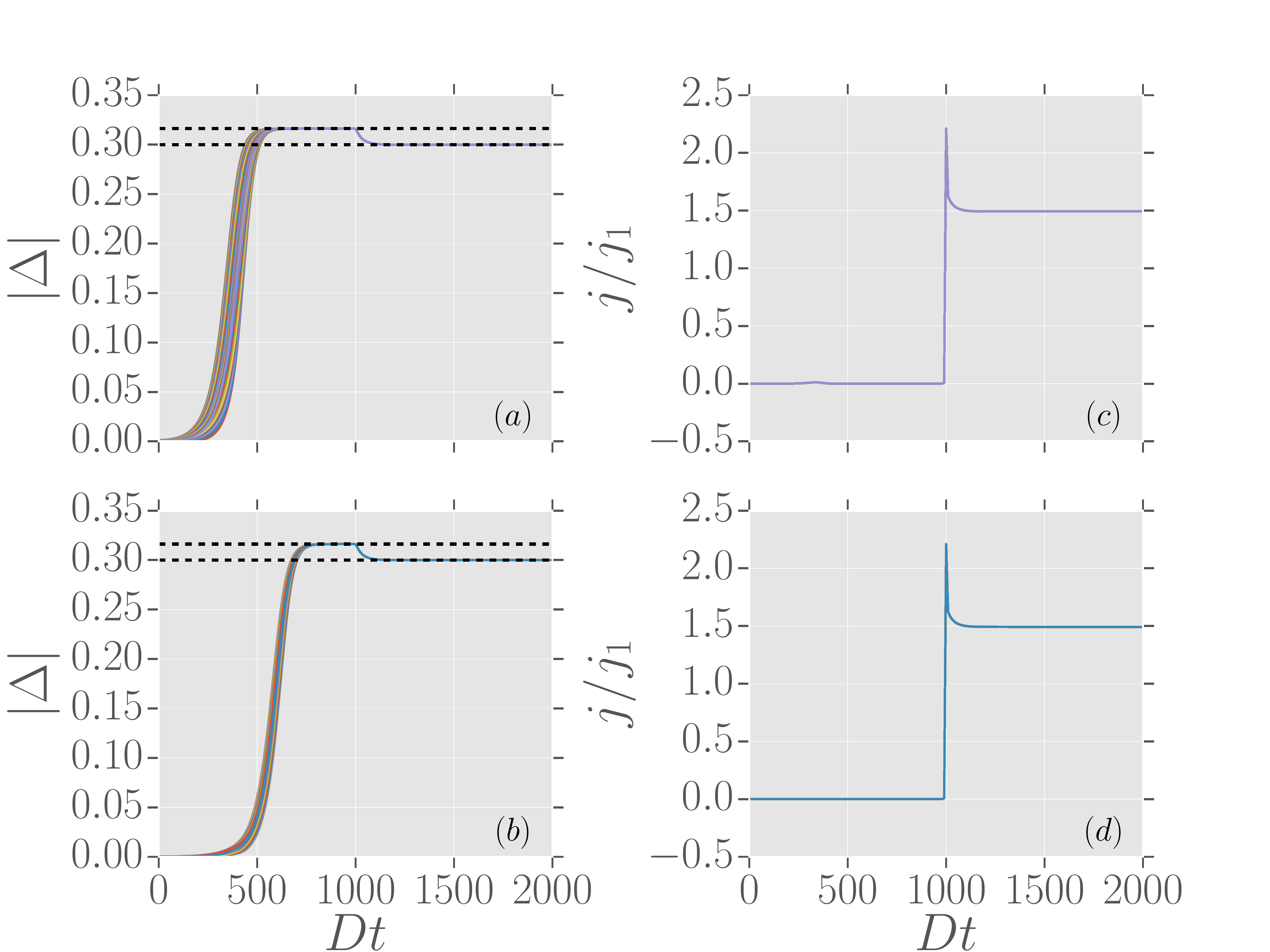}
\caption{Time evolution of order parameter magnitude (panels (a) and (b)) and x-component of current (panels (c) and (d)) computed for a one-dimensional ring (upper row, $L=100\xi$) and  a two-dimensional geometry (lower row, $L=50\xi$) for  initial configurations given by random  phases and a constant magnitude of the order parameter $\Delta_{\rm ini}=10^{-3}$. At time $Dt_p=1000$ a small electric field pulse with strength $ A\xi=0.05$ and narrow width $DT_0=3$ is applied along the ring  (upper row) or along the x-coordinate for the two-dimensional case.  At the time of the electric field pulse the phase stiffness is large enough that e the order parameter phase is not affected by the probe field. After a brief transient the sustem settles down to a state of nonzero supercurrent specified by the applied vector potential. The dashed lines in panels (a) and (b) indicate the equilibrium value of the magnitude of the order parameter as well as its value in the presence of the current induced by the electric field pulse. }  
\label{fig:plot2}
\end{center}
\end{figure}
A quench in the presence of a static vector potential (flux) may be understood in a very similar way. The minimum energy state has a phase winding  $n$ given by the nearest integer to $\Phi/\Phi_0$ and particular initial conditions may lead to long-time states characterized by a winding number $n$ which differs from this value.  If $\Phi/\Phi_0$ is not an integer the ground state will have some residual supercurrent.

The generalization of this picture to higher dimensions involves additional considerations. Random initial conditions may lead to states with vortices and antivortices (in dimension $d=2$) or vortex loops ($d=3$), whose long-time evolution involves interesting coarsening dynamics.  These issues were recently  discussed.\cite{Kobayashi16a,Kobayashi16b} Here we focus on  response to applied fields.  Fig. ~\ref{fig:plot2} compares a one and two dimensional case, showing that despite the issues of vortex/antivortex pairs the basic evolution of the gap amplitude (panels (a) and (b)) and supercurrent (panels (c) and (d)) are very similar in the two cases. We therefore believe that for the purposes of understanding the response to probe fields, consideration of the one dimensional model suffices.

\section{Response to Short Electric Field Pulse\label{sec:response}}

We now turn to the application of an electric field pulse along the wire which we describe as 
\begin{equation}
\vec{E}(t)=\vec{A}\frac{1}{T_0\cosh\left(\frac{t-t_p}{T_0}\right)^2},\label{eq:Epuls1}
\end{equation}
where $A/T_0$ is the maximal field strength, $t_p$ is the center time and $T_0$ is the width of the pulse. This electric field pulse is difficult to apply experimentally, but provides substantial physical insight (note Ref.~\onlinecite{CurrentPulse1,CurrentPulse2,CurrentPulse3}).  We will consider experimentally relevant probe pulses profiles in the next section. For simplicity we concentrate the discussion on the case where the phase winding  is zero before the pulse is applied.

The relation $\vec{E}=- \partial_t \vec{A}$ means that the E-field pulse of Eq.~\ref{eq:Epuls1}  leads to a long time increase in the vector potential $\Delta \vec{A}=\vec{A}(t\rightarrow \infty)-\vec{A}(t\rightarrow -\infty)$. Integrating Eq. ~\ref{eq:Epuls1} gives
\begin{equation}
\Delta \vec{A}(t)=-\vec{A}\frac{1+\tanh\left(\frac{t-t_p}{T_0}\right)}{2}.
\end{equation} 

\begin{figure}
\begin{center} 
\includegraphics[width=\columnwidth, angle=-0]{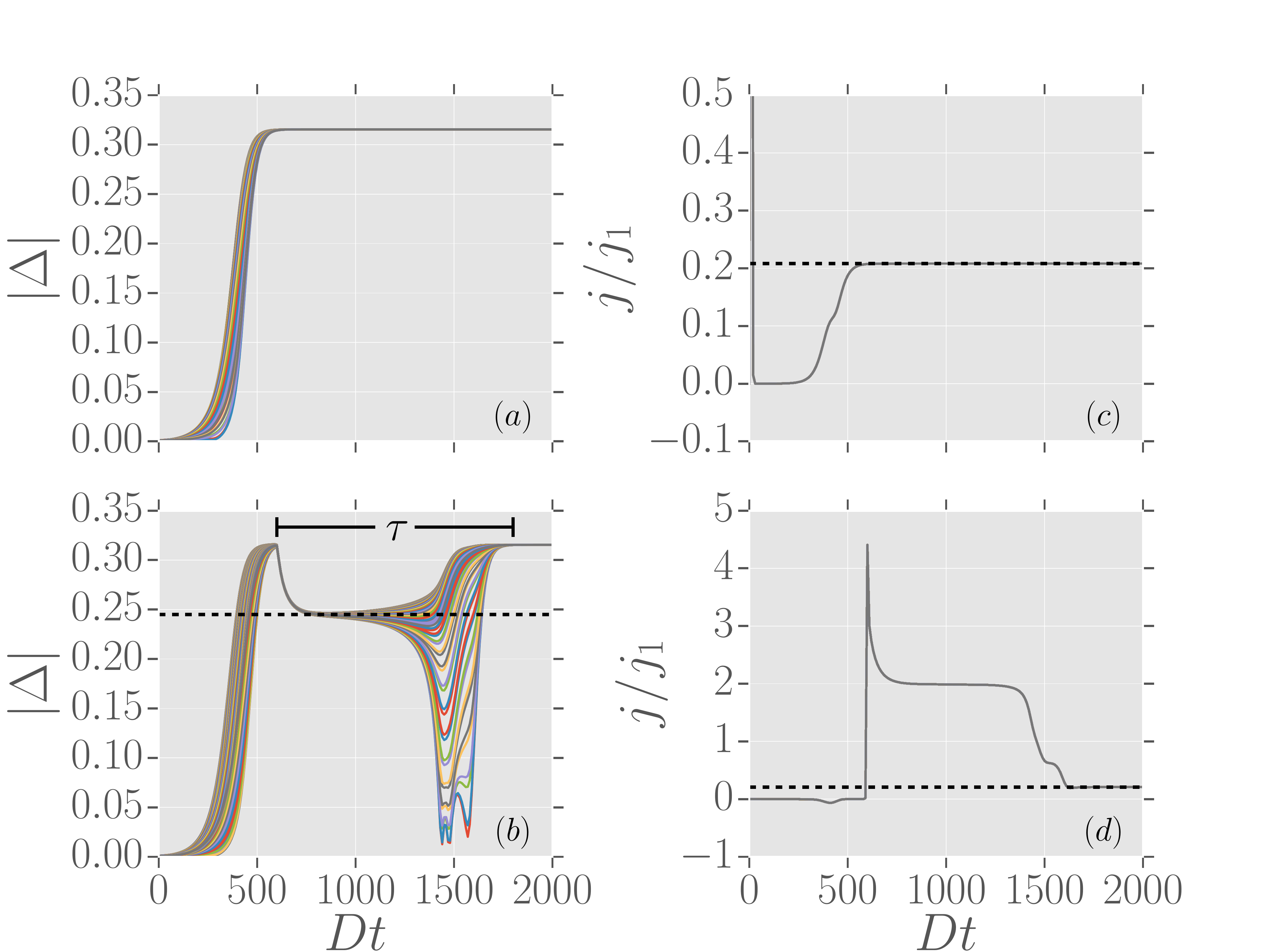}
\caption{Time evolution of order parameter magnitude (panels (a) and (b)) and  current (panels (c) and (d)) computed for a one-dimensional ring of size $L=100\xi$ subject to a quench at $t=0$ followed by an electric field pulse of magnitude $\xi A=0.1$ applied at time $Dt_p=10$ (upper panels) and $Dt_p=600$ (lower panels). The dashed line in (b) indicates the magnitude of the order parameter in the state $\nabla \phi =0$ for the given $A$. The dashed lines in (c) and (d) indicate the value of the current in the state of minimal energy corresponding to the long-time limit of the vector potential  
}  
\label{fig:plot3}
\end{center}
\end{figure}

The presence of a vector potential will lead to a supercurrent; if the current is sufficiently large, phase slips will occur, allowing the phase gradient to adapt to the vector potential  and the current to relax. The phase slip dynamics depend crucially on the magnitude of the applied field and on the timing of the pulse relative to the development of the phase stiffness of the superconducting state. If the pulse is applied at very early times, the small value of the order parameter means that phase slips are easy to drive and the phase will adapt to the vector potential, leading to minimal current at long times. On the other hand, if the pulse is applied at later times, the phase stiffness will be fully established and phase slips will only be generated if the supercurrent is sufficiently large. Fig.~\ref{fig:plot3} shows the results of applying an E-field pulse at a time $Dt_p=1000$ long after the superconducting state is established.  We see that in both the one and two dimensional cases  a supercurrent is rapidly established after the pulse and persists. The concomitant decrease of $\Delta$ is also visible.  No phase slips occur. 

For E-field pulses that are stronger, or applied earlier, phase slips may occur. The stability analysis performed in\cite{Ivlev84,Vodolazov02,Ludac08,Ludac09} reveals that within the  deterministic TDGL dynamics  for fully established phase stiffness the critical value for the vector potential is  $A_{c}=\Delta_0/(2\sqrt{3}\xi )\approx 0.091/\xi$. In Fig.~\ref{fig:plot3} we present the order parameter evolution after a quench at $t=0$ followed by  the application of a stronger electric field pulse $A=0.1/\xi>A_{c}$.  The upper panels shows the results when the pulse is applied the early time,  $Dt_p=10$.  Comparison of panel (a) to the corresponding panel of  Fig.~\ref{fig:plot2} shows that the time evolution of the magnitude is almost unaffected by the change in the vector potential:  the phase simply rearranges to compensate the external vector potential to the extent possible given the quantization of the phase winding.  The evolution of the current reveals similar physics: we see that after an initial current pulse visible as the vertical region adjacent to the $t=0$ axis, the current is very small (corresponding to the very small order parameter amplitude) and then increases as the order parameter increases, eventually saturating at the value given by optimal winding number.  

The lower panels of Fig.~\ref{fig:plot3} show results when the pulse is applied at a late time, $Dt_p=600$, after the superconducting state is almost fully established.  In this case the initial response of the system is to keep the phase fixed ($\nabla\phi=0$), to reduce the order parameter to the value (dashed line) corresponding to  $\nabla \phi=0$ and the given $A$, and to build up a supercurrent $\sim \Delta^2 A$. We see however that this large-current state is only an intermediate time regime: at a longer time phase slip events  occur that relax the system back to the minimum energy state, with minimal current and maximal order parameter amplitude.

\begin{figure}
\begin{center} 
\includegraphics[width=\columnwidth, angle=-0]{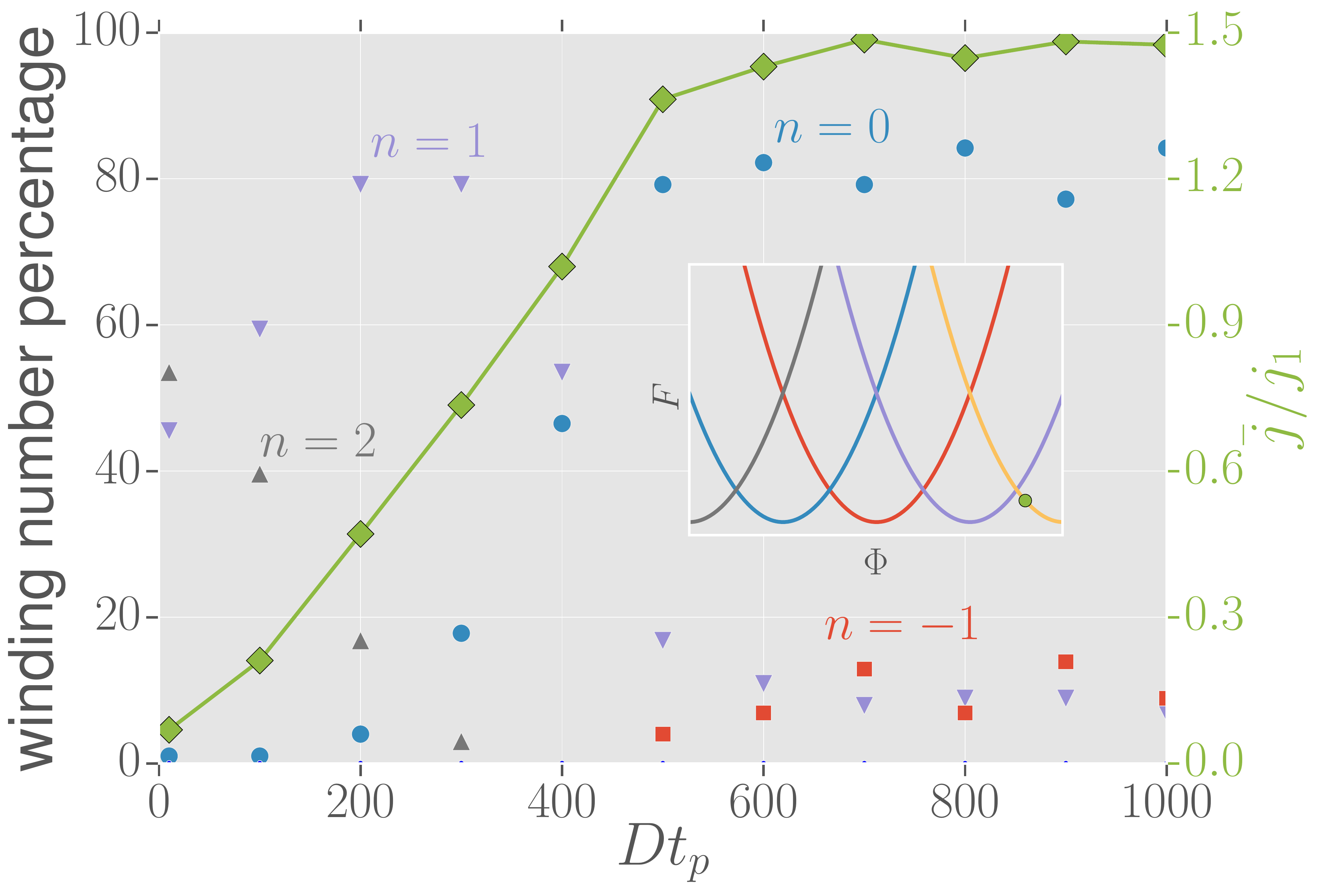}
\caption{Characterization of long -time ($Dt=2000$) limit of states produced for a system  of length $L=100\xi$ by time evolution of 100 random initial conditions  (random initial phases and  constant magnitude of the order parameter $\Delta_{\rm ini}=10^{-3}$ ) subject to an electric field pulse of the form of Eq. ~\ref{eq:Epuls1} (strength $\xi A=0.05$, width $DT_0=3$ and center time $t_p$). Filled diamonds connected by line (green on-line): long time limit of current averaged over all 100 initial conditions (right axis). Squares, circles, up and down triangles indicate percentage of initial conditions that produce winding number of $n=-2,~-1,~0,~1$ respectively. }
\label{fig:cross}
\end{center}
\end{figure}

To understand the degree to which phase slips occur we  computed the long-time limit of the winding number and current following from 100 randomly chosen initial conditions, for electric field pulses applied at a series of times ranging from very early, when the magnitude of the superconducting order parameter is negligibly small, to late, when the superfluidity  is well established. We chose a pulse strength $\xi A=0.05$ and a system of length $L=100\xi$, so the flux $\Phi=\frac{5}{\pi}\Phi_0$. This choice of flux puts the system very close to the crossing point of the parabolas (as indicated in the inset of Fig.~\ref{fig:cross}).  For each initial condition we computed the long time limit of the winding number and the current. The diamonds connected by a solid line in Fig. ~\ref{fig:cross} (green on-line, numerical values given on the right axis) show  the long time limit of the current $\bar j$, averaged over all 100 initial conditions  and plotted as a function of the time the E-field pulse was applied. A pulse applied at an early time leads to a small current; a pulse applied at late times leads to a large current. 

We have also analysed the statistics of phase slip events. The squares, circles, down triangles and up triangles show the percentage of initial conditions (left axis) leading to states with winding number $n=-1,0,1,2$ respectively for a probe field applied at the given time. If the electric field pulse is applied early  we find that the phase conforms  to the vector potential, whose long time limit corresponds to a flux $\Phi=\frac{5}{\pi}\Phi_0\approx 1.59\Phi_0$   yielding a high percentage of states with $n=2$ and $n=1$ phase winding.  As the time of application of the electric field pulse is increased,  the phase stiffness increases  and the likelihood of phase slips (needed to reach the $n\neq 0$ states) go down. We see that the probability of finding a long time state with winding number $n=2$ rapidly decreases; and as the time of application of the pulse further increases the vast majority of the states have $n=0$ with a small probability of $n=\pm 1$.

\begin{figure*}
\begin{center} 
\includegraphics[width=2\columnwidth, angle=-0]{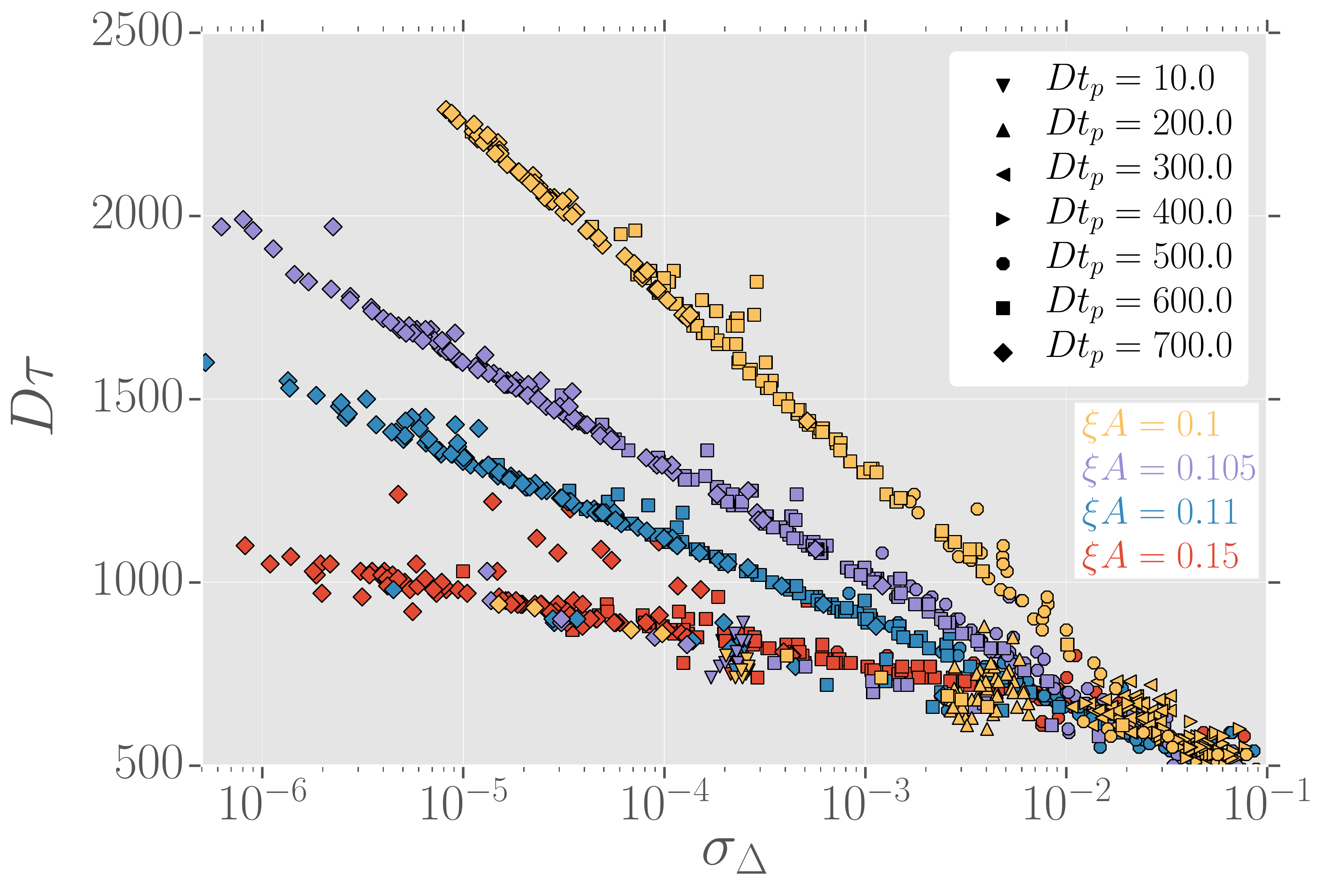}
\caption{Width $\tau$ of the current plateau (time scale on which the phase winding rearranges due to phase slips) as defined in the lower row of Fig.~\ref{fig:plot3} found following an electric field pulse with  $DT_0=3$ and different field strengths, applied at different times (see legends) and plotted agains  the standard deviation in the magnitude of the order parameter $\sigma_\Delta$ comuted at the time of the pulse.   
}  
\label{fig:Fluc}
\end{center}
\end{figure*}

\begin{figure}
\begin{center} 
\includegraphics[width=\columnwidth, angle=-0]{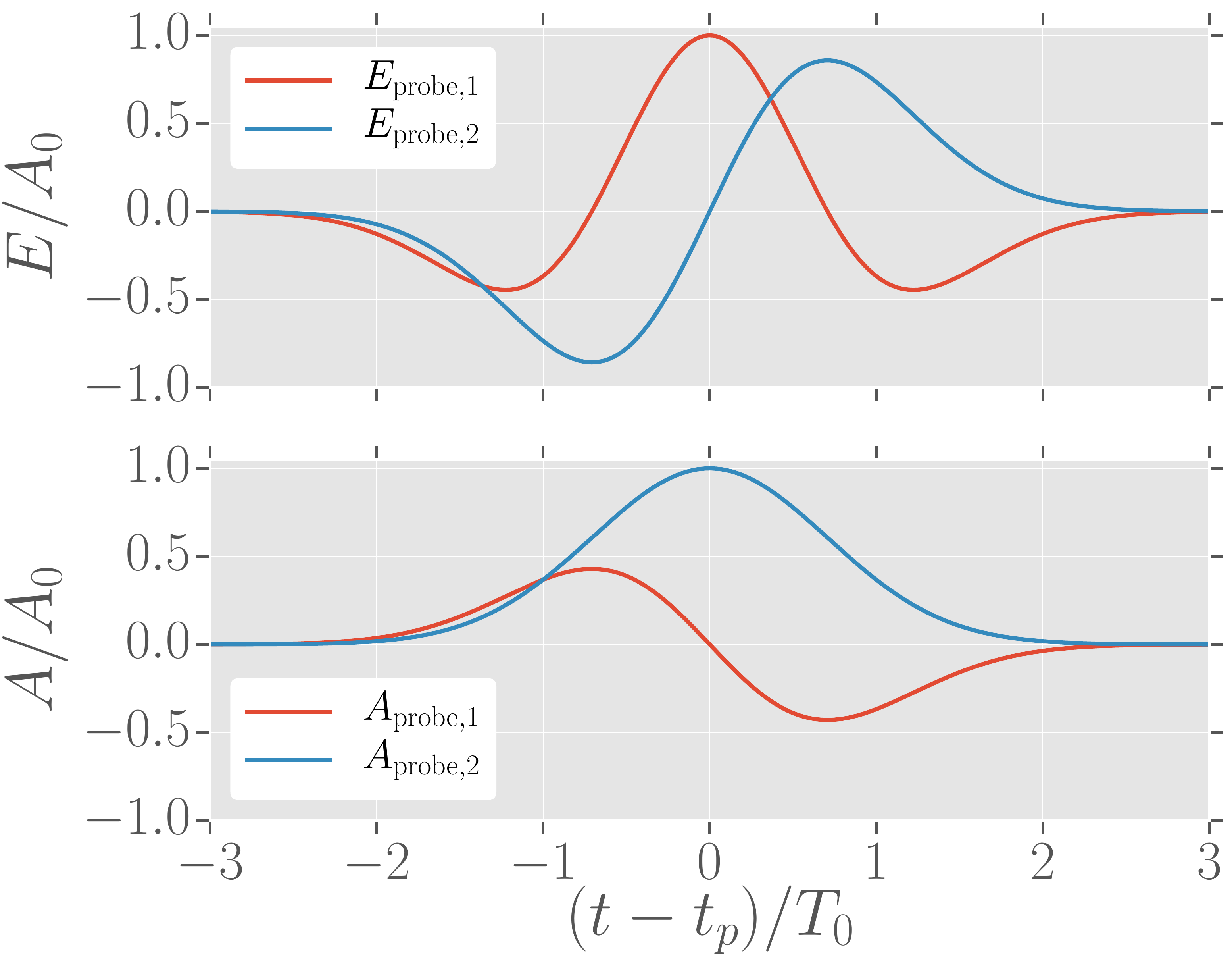}
\caption{Top panel: Electric field profile for probe functions given by the functional form of the vector potentials $A_{\rm probe, 1}(t)$ or $A_{\rm probe, 2}(t)$ with $A(t)=-\partial_t E(t)$. Bottom panel: corresponding vector potential.  }  
\label{fig:funcft}
\end{center}
\end{figure}

Finally, we turn our attention to the lifetime of the supercurrent plateau obtained for stronger electric field pulses. If  $A>A_{c}$, dynamic phase slips become activated leading to transitions to lower-current, energetically favored states. For $A>A_c$ the typical situation is shown in panel (d) of Fig.~\ref{fig:plot3}: we have a supercurrent plateau that lasts for a time $\tau$ before a set of phase slip events occur that cause the system to transition back to a state of zero winding number and vanishing current. We calculated  $\tau$ of the supercurrent (as defined in Fig.~\ref{fig:plot3})  for a total of $100$ initial conditions $A$  and different center times $Dt_p$ of the electric field pulse.   Fig.~\ref{fig:Fluc}  shows that  $\tau$ has a systematic dependence on the strength of the applied field. Stronger E-field pulses leave larger vector potentials, which more easily drive phase slips. We also see that if the pulse is applied at relatively early times $Dt_p\lesssim 400$ the waiting time for a phase slip is relatively short $D\tau \lesssim 700$ the systematics are less clear. In this regime the phase stiffness is small enough that phase slips occur relatively easily. For pulses applied at late times, we also see a systematic dependence on the magnitude of the order parameter fluctuations. These fluctuations are parametrized by $\sigma_\Delta$, the root mean square order parameter fluctuation, computed  at the  time at which the pulse is applied.  The dependence on $\sigma_\Delta$ is logarithmic.   The outliers to this logarithmic dependency are cases either where  phase stiffness is not yet established  at the time of the electric field pulse (approximately $Dt_p<500$) or if the winding number before the application of the electric field pulse happens to be $n> 0$.
\begin{figure}
\begin{center} 
\includegraphics[width=\columnwidth, angle=-0]{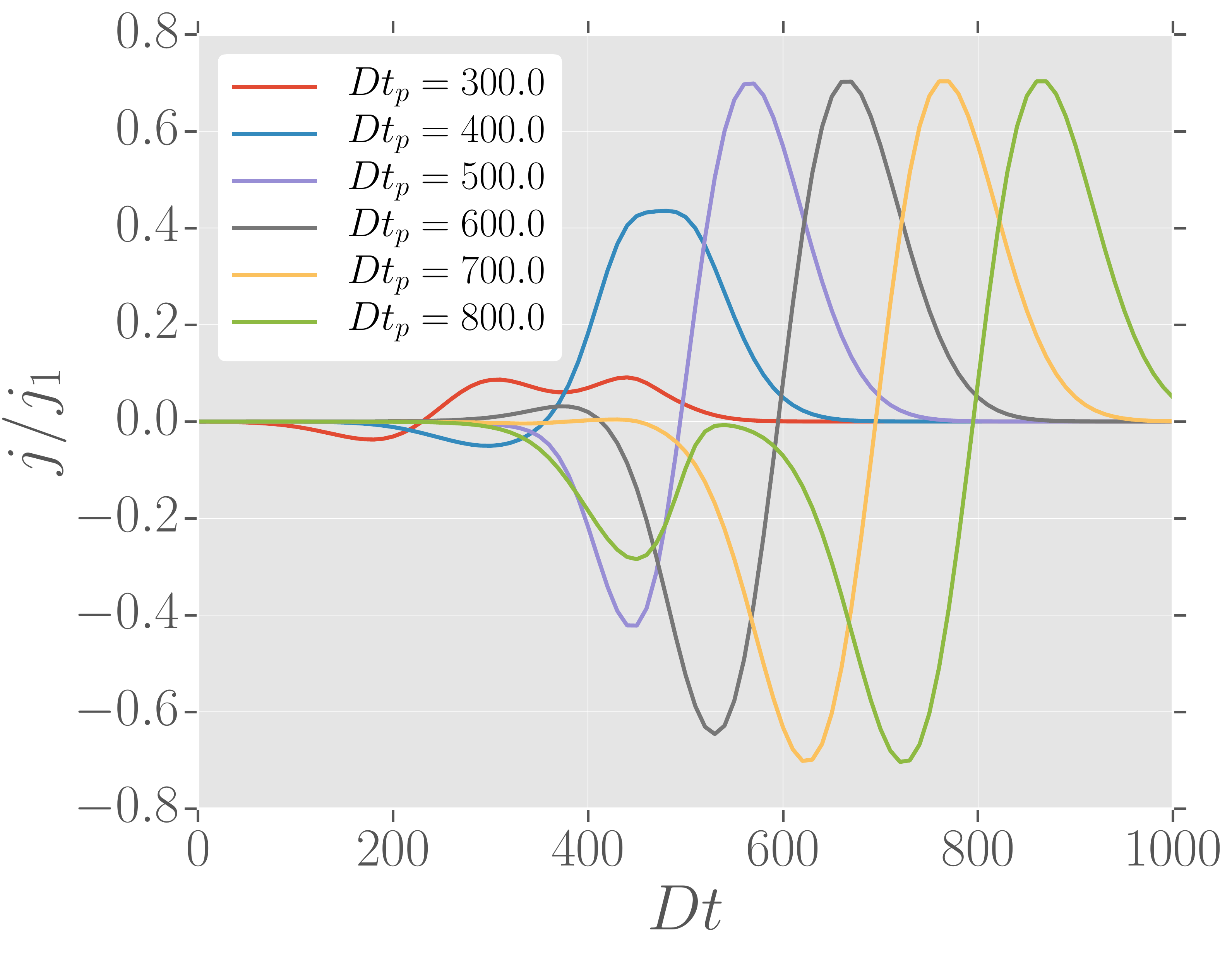}
\caption{Typical response of the current (lines) to a type-I electric field pulse with width $DT_0=100$  applied at different center times $t_p$. The other parameters are $L=100$, $\xi A_0=0.05$ and $\sigma_{\rm ini}=10^{-3}$. The other parameters are $L=100$, $\xi A_0=0.05$ and $\sigma_{\rm ini}=10^{-3}$. }  
\label{fig:Resp1}
\end{center}
\end{figure}

\section{Response to Physical Electric Field Pulse \label{sec:PhysicalPulse}}
Many time-domain experiments,\cite{Averitt00,Larsen11,Matsunaga12a}  utilize an electric field probe pulse of approximately the form  (we also give the corresponding vector potential)
\begin{align}
E_{\rm probe, 1}(t)&=A_0\left(1-2\frac{(t-t_p)^2}{T_0^2}\right)e^{-(t-t_p)^2/T_0^2}\\
A_{\rm probe, 1}(t)&=-A_0\frac{t-t_p}{T_0}e^{-(t-t_p)^2/T_0^2},
\end{align}
with the peak electric field $E_0=A_0/T_0$.

In Ref.~\onlinecite{Kennes17} we proposed a second form of the probe field 
\begin{align}
E_{\rm probe, 2}(t)&=2A_0\frac{(t-t_p)}{T_0}e^{-(t-t_p)^2/T_0^2}\\
A_{\rm probe, 2}(t)&=A_0e^{-(t-t_p)^2/T_0^2},
\end{align}
These functional forms are depicted in Fig.~\ref{fig:funcft}. In the following we will denote them as type-I and type-II probe pulses, respectively.  The form $E_1$ is more convenient experimentally, but $E_2$ allows for the reconstruction of the time dependent superfluid stiffness via a time integral of the induced current. The parameter $a$ tunes the width of the probe pulses.

\begin{figure}
\begin{center} 
\includegraphics[width=\columnwidth, angle=-0]{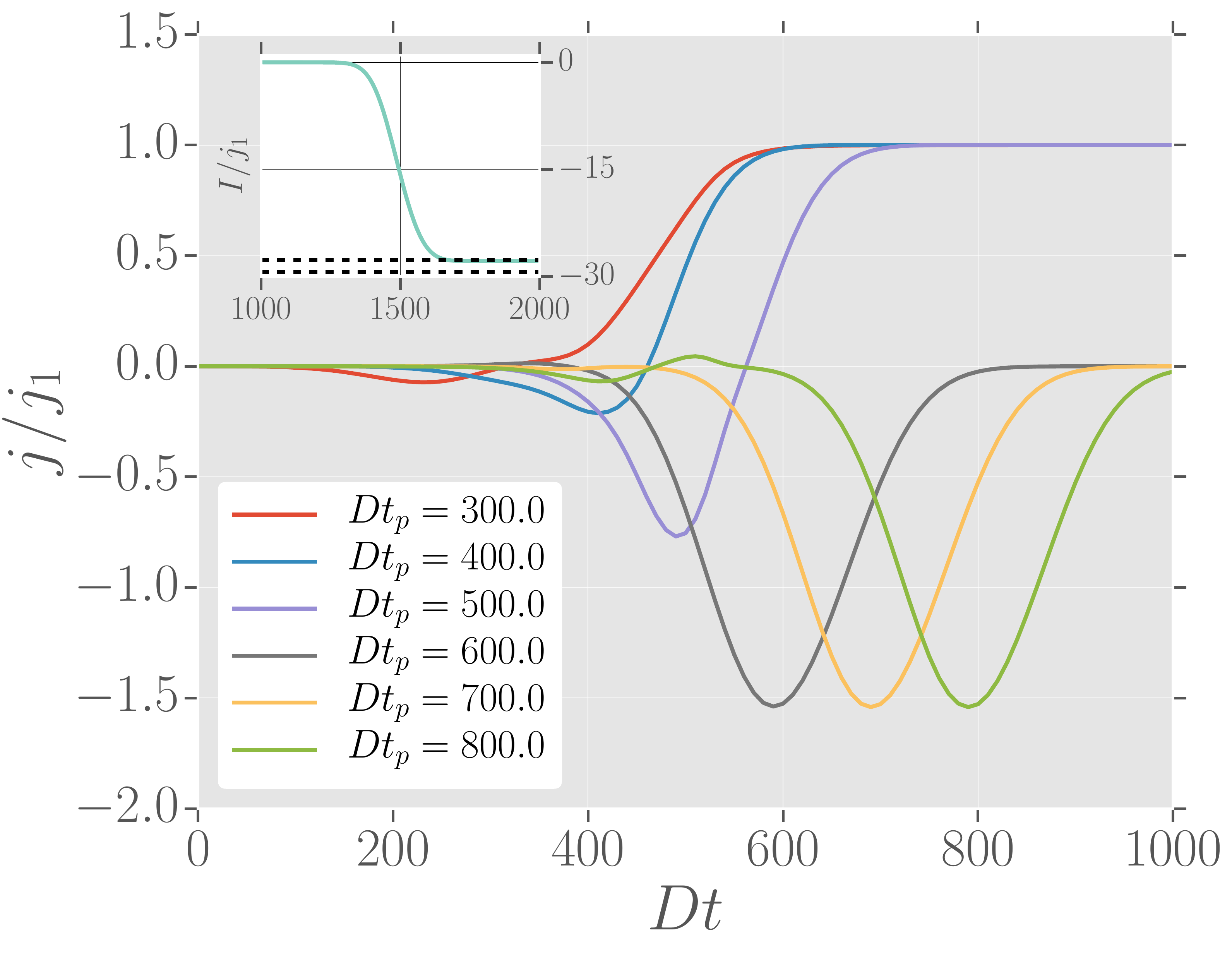}
\caption{Main panel: Typical response of the current (lines) to a type-II electric field pulse with width $DT_0=100$  applied at different center times $t_p$. The other parameters are $L=100$, $\xi A_0=0.05$ and $\sigma_{\rm ini}=10^{-3}$. Inset: integrated current $I(t)$ Eq.~\eqref{eq:I} from times $t_{\rm l}=1000$ for a very late probe pulse $t_p=1500$. Dashed lines are the full Eq.~\eqref{eq:Iinf} (upper dashed line) and keeping only the linear order in \eqref{eq:Iinf} (lower dashed line), respectively.
}  
\label{fig:Resp2}
\end{center}
\end{figure}

Fig.~\ref{fig:Resp1} and fig.~\ref{fig:Resp2} summarize the current response to a type-I and type-II electric field probe pulse, respectively. If the phase is not stiff the current response to an electric field probe is very weak. As the phase stiffens the response's line shape increasingly starts to corresponds to the input vector potential until at large probe times $t_p$ the current perfectly follows the potential change as expected from $j\sim A(t)$ at $\nabla \phi=0$. In Ref.~\onlinecite{Kennes17} we argue that assuming the phase is stiff the integrated current 
\begin{equation}
I(t)=\int\limits_{t_{\rm l}}^td\tau j(\tau)\label{eq:I}
\end{equation} 
resulting from a type-II probe pulse can be linked to the superfluid stiffness. Here we introduce a lower cutoff $t_{\rm l}$ such that we do not integrate over the current resulting out of the quench dynamics. Thus $t_{\rm l}$ must be larger then the time needed to build up the phase stiffness (for the paramter used here $Dt_{\rm l}\approx 500$) to relate the integrated current to the superfluid stiffness. At large times where the phase is indeed stiff the integrated current should reach the asymptotic value
\begin{equation}
I(\infty)=-2\sigma\tau_s\xi A\sqrt{\pi T_0}\left[\Delta_0^2-\frac{4}{\sqrt{3}}(\xi A)^2\right].\label{eq:Iinf}
\end{equation}  
We include here the $A^3$ correction resulting from the $\Delta$ dependence on the  supercurrent, which was neglected in  Ref.~\onlinecite{Kennes17}.

\begin{figure}
\begin{center} 
\includegraphics[width=\columnwidth, angle=-0]{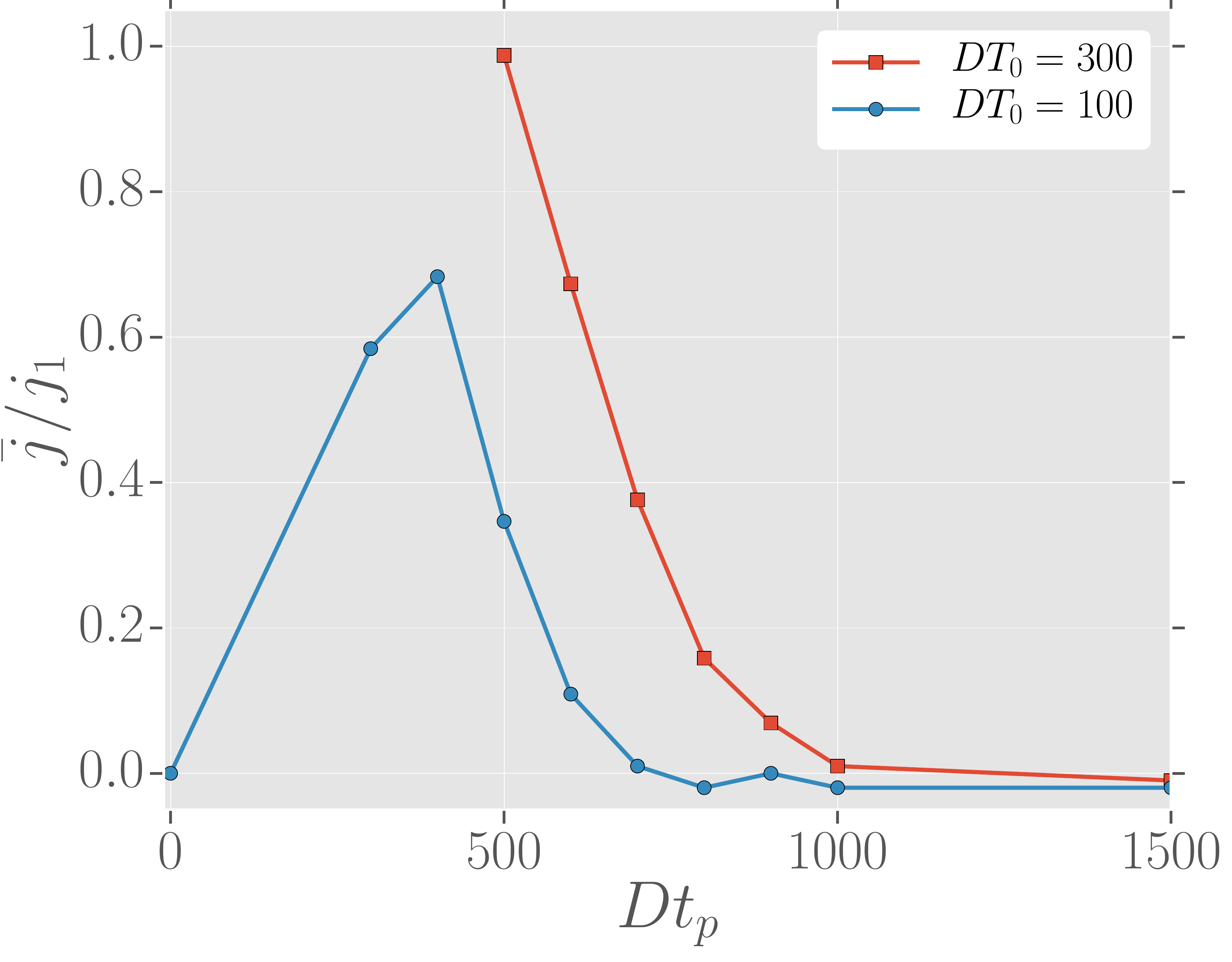}
\caption{ Average current response $\bar j$ (symbols) sampled over 100 initial conditions to a type-II electric probe field. The other paraemters are the same as in Fig.~\ref{fig:Resp2}. Lines are guides to the eyes only.}  
\label{fig:Resp22}
\end{center}
\end{figure}

The integrated current $I(t)$ for a type-II probe field applied at large times  $Dt_p=1500$ is shown in the inset of Fig.~\ref{fig:Resp2}. We choose $Dt_{\rm l}=1000$ to cutoff the small contribution to the current arising due to the quench dynamics. As the upper dashed line we give the value predicted from Eq.~\eqref{eq:Iinf} and as the lower dashed line the same but keeping only the linear order in $A$. The former agrees perfectly, while the later shows small deviations as expected for the shown small value of $\xi A_0=0.05$. 

We note an interesting feature in the current response. While for both pulse types, the time integral of the electric field vanishes at long times (so no long-time vector potential is induced), the time dependence of the superconductivity means that the integral of the  product of the electric field and the superfluid  response need not vanish. In physical terms, in a nonequilibrium situation the supercurrent created during the second part of the electric field cycle need not cancel supercurrent created in the first part, leaving a net supercurrent even for a type I or II pulse. The effect is most pronounced for a relatively wide type II pulse with center time corresponding to the time over which the phase stiffness is becoming established. In this case the negative half of the electric field pulse might be in the regime where the phase is not stiff and  a negligible supercurrent is induced while for the positive half of the electric field pulse  the phase is stiff and a current is induced.  To analyze this in more depth we present in Fig.~\ref{fig:Resp22} the long time current $\bar j$ averaged over $100$ initial conditions for the same parameters as in Fig.~\ref{fig:Resp2} and two widths of the type-II electric field probe pulse $T_0=100$ and $T_0=300$. As the center time $t_p$ crosses through the build up time of the superconductor ($Dt\approx500$) the current first rises up and then returns to zero, in accordance to the picture drawn above.

\section{Conclusion\label{sec:conclusion}}

In this paper we have used the time dependent Ginzburg-Landau equations to  address the response to a probe field of a system quenched into a superconducting state. The key issue is that a superfluid response requires a non-negligible phase stiffness, which has to build up dynamically after a quench. The value of the phase stiffness controls the likelihood of phase slips and the magnitude of the electromagnetic response. An important related issue is that the quench dynamics correspond to evolution in time from random initial conditions; these lead to a distribution of states, some of which are  metastable states carrying a supercurrent even in the absence of externally applied fields. Such states have a different electromagnetic response than do non current-carrying states. A third important point  is whether the probe is strong enough to activate phase slip processes. If the probe is applied before the superconducting order is established, phase slips occur so that the vector potential is compensated and no long term supercurrent is induced, whereas for a weaker probe pulse a state with a long-lived supercurrent is established.  We find that the lifetime of a state with a long-lived supercurrent scales logarithmically with the amplitude of fluctuations in the magnitude of the order parameter. 

We considered the response to two types of probe line-shapes. One is the electric field pulse, in which the time integral of the field is non-zero so the vector potential is different at long times after the pulse than it was before the pulse. In the other (of which we considered two variants) the time integral of the electric field vanishes, so no vector potential is left at long times. The response to an electric field pulse provides useful physical insights, but pulses in which the time integral of the electric field vanishes are  more easily achievable in experiments.

We studied the changes in system response as the time of probe application is varied relativel to the time of the onset of superconducting phase stiffness. For the case where the probe is applied before the superconducting order builds up, we find that the phase of the superconducting order parameter adopts to the vector potential change and there is no clear superconducting response. In the opposite case where the probe is applied after phase stiffness is established, we find a canonical superconducting response.  In the intermediate regime, where the center time of the probe pulse aligns with the build up time of the superconducting phase stiffness, we report that an asymptotic supercurrent can be induced at large times although the total vector potential does not change overall. This is because the one cycle of the electric field pulse applied at times where the phase is not stiff and can thus be compensated by the phase, while during the times of the second cycle the phase is stiff and thus the system will respond with the build up of a supercurrent. If that current is too small to activate the phase slip process it will have infinite lifetime.   

Several generalizations of our work would be of interest. Extending our analysis, which is based on simple theoretical models, to more realistic situations such as one dimensional wires with a finite transverse dimension, is important because the wire thickness will affect the energetics and dynamics of phase slips. Further consideration of experiments that might reveal the phase slip dynamics is important.  The mesoscopic considerations of disorder and sample to sample fluctuations are also of interest. Further, our analysis of higher dimensional situations was limited. Our work indicates that the basic issues relating to the timing of the probe field relative to the onset of phase stiffness are not strongly dimension dependent, but the interplay of the timing of probe fields with other aspects of the quench physics including structure of vortex-antivortex pairs or vortex loops requires further analysis. Additionally, treating thermal fluctuations and the pinning of phase slip center by impurities is a fascinating and important avenue of future research.

{\it Acknowledgements:} AJM was supported by the Basic Energy Sciences Program of the US Department of Energy under grant \text{DE-SC}0012592. DMK was supported by DFG KE 2115/1-1. Simulations were performed with computing resources granted by RWTH Aachen University under project rwth0013.  

\appendix
\setcounter{equation}{0}
\section{Numerical Procedure}
\label{Appendix:a}
In this appendix we present some details of our numerical procedures. 

\begin{figure}
\begin{center} 
\includegraphics[width=\columnwidth, angle=-0]{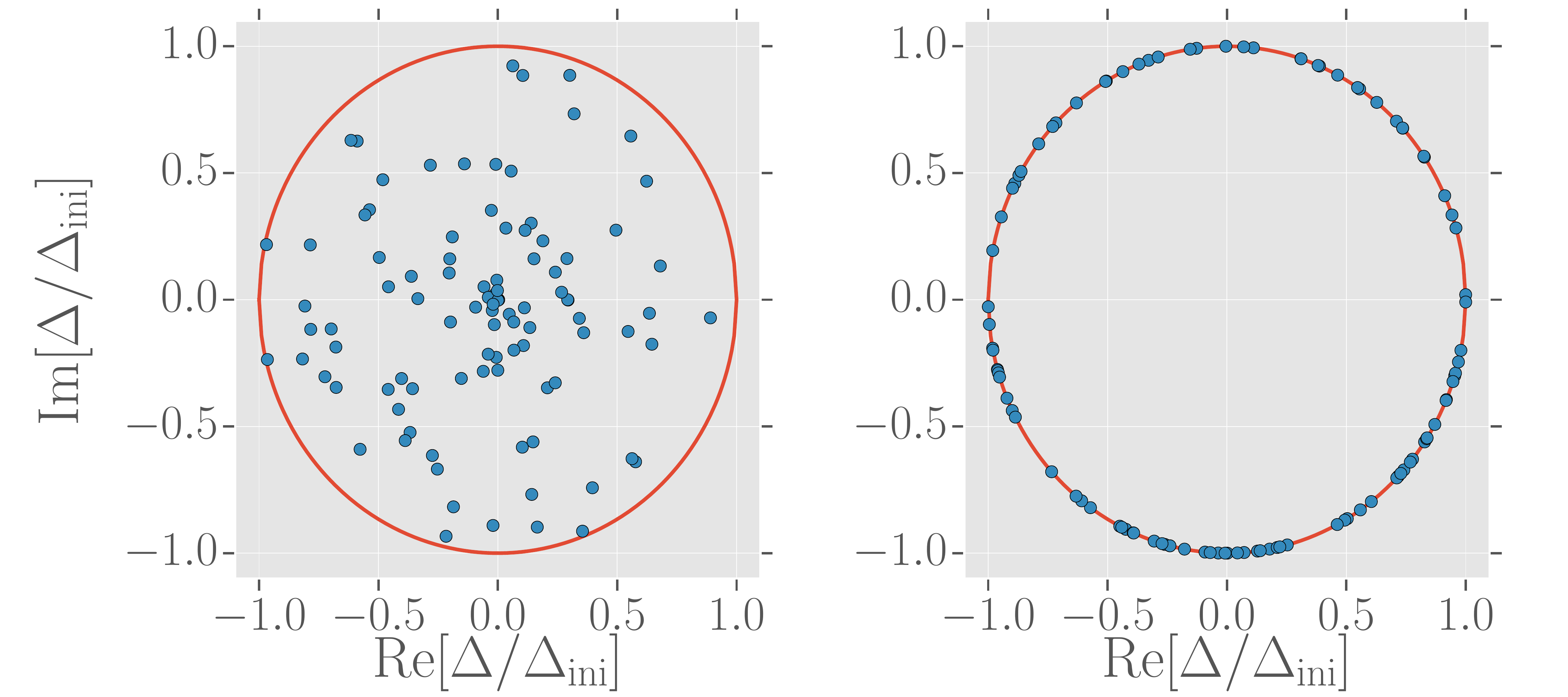}
\caption{Representation of initial values (filled circles, blue on-line) of magnitude and phase of superconducting order parameter on each of the 100 sites of a $L=100$ site  computational lattice for the  two cases considered in this paper. Left Panel: one particular set of initial conditions for the case in which the magnitude of the order parameter on each site is chosen randomly from the a uniform distribution extending from $|\Delta|\in [0,\Delta_{\rm ini})$ while the  phase of the order parameter is chosen randomly from a distribution uniform over the interval $0-2\pi$. Right Panel: one particular set of initial conditions for the case in which the  magnitude of the order parameter taken to be constant ($|\Delta|=\Delta_{\rm ini}$), but the  phase is chosen randomly from a distribution uniform over the interval $0-2\pi$. The solid line (red online) shows the case $|\Delta|=\Delta_{\rm ini}$. 
}  
\label{fig:ini_cond_1}
\end{center}
\end{figure}

\begin{figure}
\begin{center} 
\includegraphics[width=\columnwidth, angle=-0]{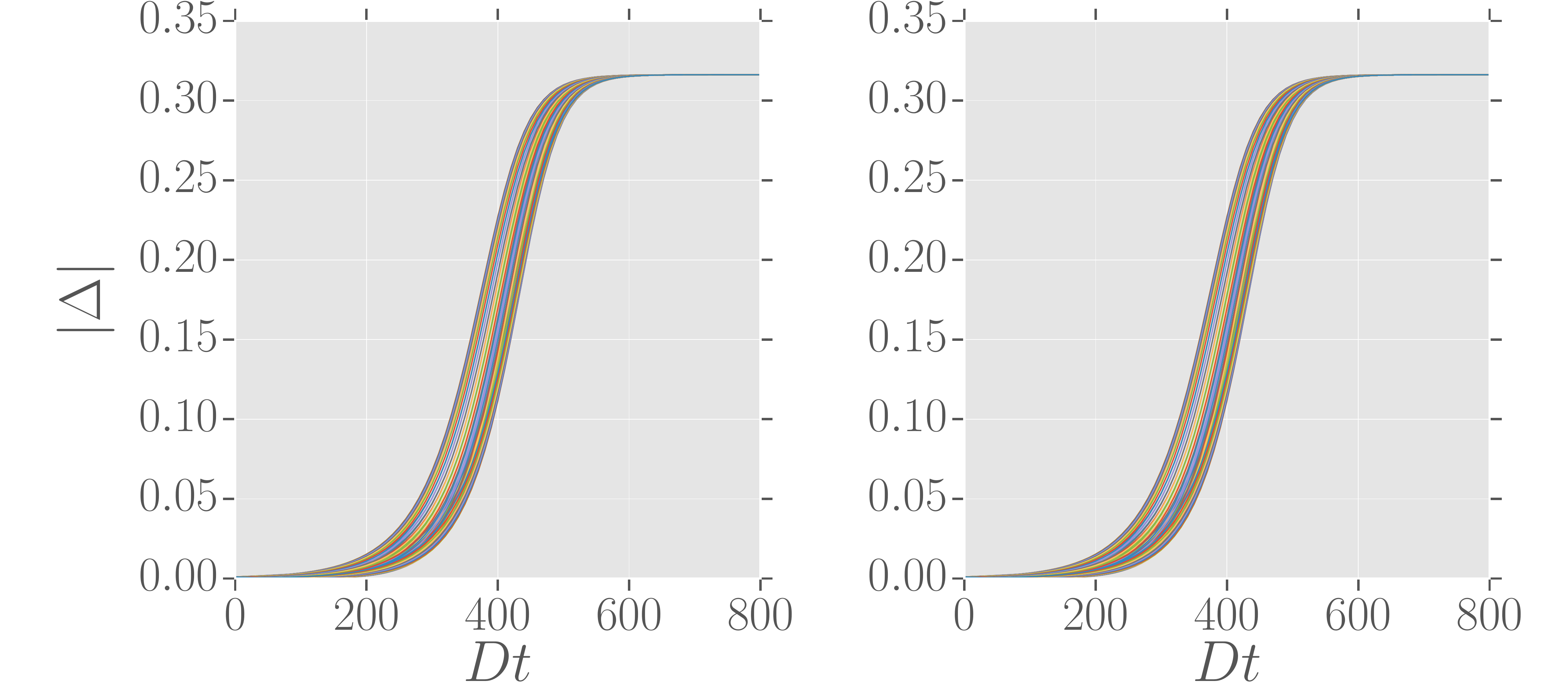}
\caption{Time evolution of the order parameter magnitude  for a ring of size $L=100\xi$  obtained  by time-evolving an  initial configuration specified by random  phases  and a random magnitude of the order 
$\left|\Delta\right|\in [0,\Delta_{\rm ini})$ on each site  (left panel) and by time-evolving an initial configuration specified  by random  phases on each site and constant magnitude of the order parameter $\left| \Delta\right|=\Delta_{\rm ini}$ (right panel) with $\left|\Delta_{\rm ini}\right|=10^{-3}$. The two initial conditions  give equivalent results.}  
\label{fig:plot02}
\end{center}
\end{figure}

A quench involves an evolution forward in time from initial conditions that are determined by the fluctuations in the pre-quench state. In the superconducting case of interest here one has a two-component order parameter, which may be characterized by a magnitude and a phase or by the amplitudes of the real and imaginary parts. The actual initial conditions involve a  distribution of both magnitude and phase (or equivalently of real and imaginary parts) which is random in space and time but correlated over length scales of the order of a bare coherence length $\xi$. Since the basic length scale in our theory and in our numerics is $\xi$ we simply take the order parameter to be random from site to site of our computational lattice. One may also as a simplification in the numerical consider initial conditions in which the magnitude of the order parameter is fixed and only the phase varies from site to site. The two choices of initial condition are represented in Fig. ~\ref{fig:ini_cond_1}. We have found that the two initial conditions lead to equivalent results, which is shown in Fig.~\ref{fig:plot02}.

\begin{figure}
\begin{center} 
\includegraphics[width=\columnwidth, angle=-0]{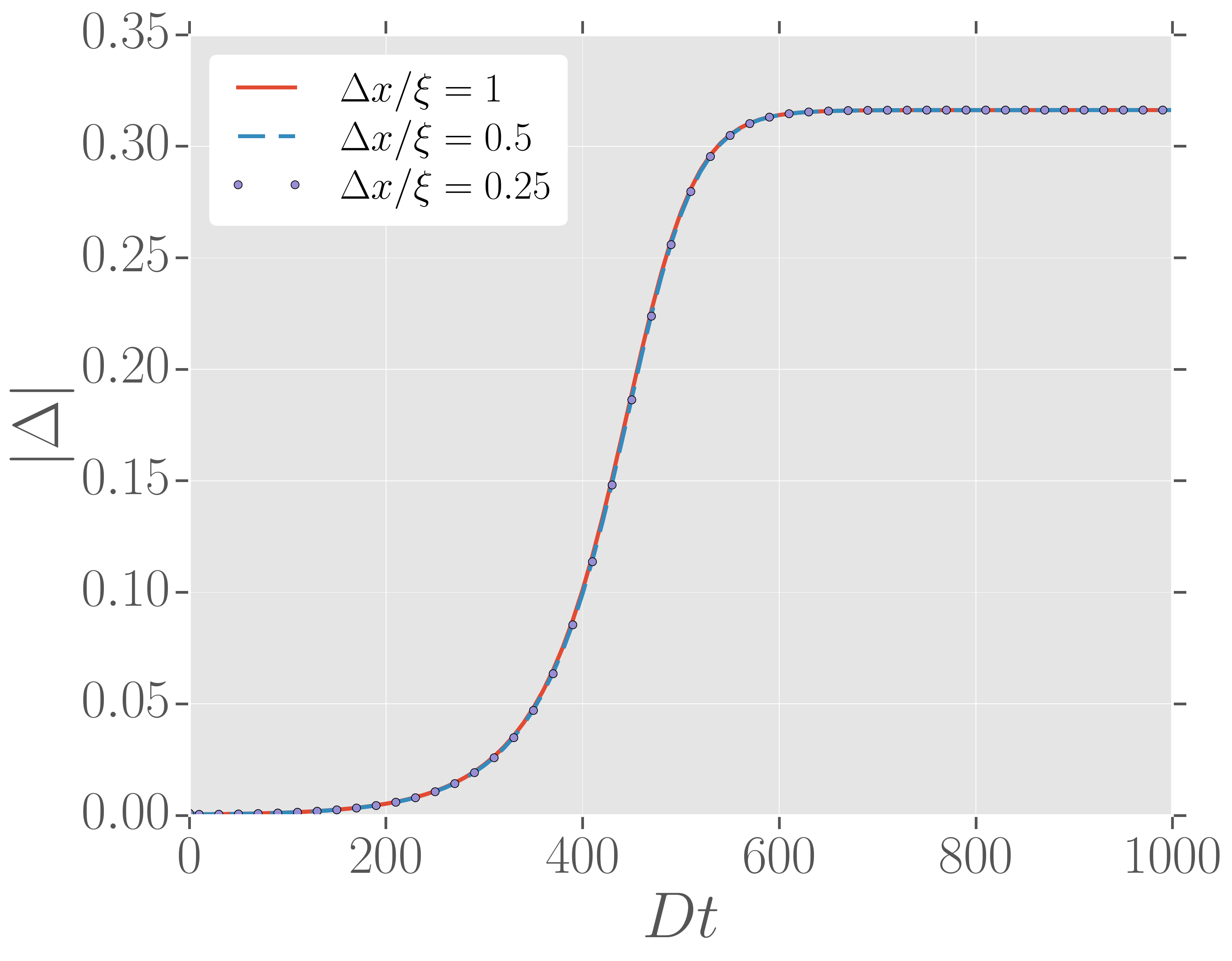}
\caption{Time dependence of the magnitude of the order parameter at one site for different $\Delta x$ and mutually the same initial conditions. The parameters are the same as in Fig.~\ref{fig:plot1}.
}  
\label{fig:conv}
\end{center}
\end{figure}

We solve the TDGL equations by integrating forward in time using a first order Euler method. The algorithm requires a step size $\Delta x$ in space and $\Delta t$ in time. We have found  that choosing $\Delta x=\xi$ suffices. To verify this we consider how  the results shown in Fig.~\ref{fig:plot1} change as the step size is reduced.  We compare numerically three values of $\Delta x/\xi=1,0.5,0.25$ keeping $L/\xi=100$ constant. This means that we define lattices of different number of lattice sites  $100$, $200$ and $400$, respectively. To address the numerical convergence we need to choose the same initial conditions for all three values of $\Delta x$. To do so, we draw a random initial condition for our largest value of $\Delta x/\xi=1$ and then for the smaller values of $\Delta x$ copy the each value of the initial condition onto a pair of sites, for $\Delta x/\xi =0.5$, or a quadruple of sites, for $\Delta x/\xi =0.25$. The results are summarized in Fig.~\ref{fig:conv} showing the magnitude of the order parameter for a typical initial condition at one of the lattice sites. We see that decreasing the step size has no effect on the results

Similarly we find that comparing $D \Delta t=0.01$ to $D \Delta t=0.001$ gives converged results and we choose the ladder.

{}

\end{document}